\documentclass[runningheads]{llncs}
\usepackage{graphicx}
\usepackage{url}
\usepackage{multirow}
\usepackage{booktabs}
\makeatletter
\newcommand{\printfnsymbol}[1]{%
  \textsuperscript{\@fnsymbol{#1}}%
}
\makeatother
\begin{document}
\title{Network Dependency Index Stratified Subnetwork Analysis of Functional Connectomes: An application to Autism}
\titlerunning{Network Dependency Index Subnetwork Analysis of Functional Connectomes}
%
\author{Ai Wern Chung\inst{1}\orcidID{0000-0002-0905-6698} \and
Markus D. Schirmer\inst{2,3}\orcidID{0000-0001-9561-0239}}
\authorrunning{Chung and Schirmer}
%
\institute{Fetal-Neonatal Neuroimaging \& Developmental Science Center, Boston Children’s Hospital, Harvard Medical School, Boston, MA, USA  \and
Stroke Division \& Massachusetts General Hospital, J. Philip Kistler Stroke Research Center, Harvard Medical School, Boston, USA \and
Department of Population Health Sciences, German Centre for Neurodegenerative Diseases (DZNE), Germany \email{mschirmer1@mgh.harvard.edu}}

\maketitle              

\begin{abstract}
Autism spectrum disorder (ASD) is a neurodevelopmental condition impacting high-level cognitive processing and social behavior. Recognizing the distributed nature of brain function, neuroscientists are exploiting the connectome to aid with the characterization of this complex disease. The human connectome has demonstrated the brain to be a highly organized system with a centralized core vital for effective function. As such, many have used this topological principle to not only assess core regions, but have stratified the remaining graph into subnetworks depending on their relation to the core. Subnetworks are then utilized to further understand the supporting role of more peripheral nodes with respects to the overall function in the network. A recently proposed framework for subnetwork definition is based on the network dependency index (NDI), a measure of a node's importance based on its contribution to overall efficiency in the network, and the derived subnetworks, or Tiers, have been shown to be largely stable across ages in structural networks. Here, we extend the NDI framework to test its efficacy against a number experimental conditions. We first not only demonstrated NDI's feasibility on resting-state functional MRI data, but also its stability irrespective of the group connectome on which NDI was determined for various edge thresholds. Secondly, by comparing network theory measures of transitivity and efficiency, significant group differences were identified in NDI Tiers of greatest importance. This demonstrates the efficacy of utilizing NDI stratified subnetworks, which can help to improve our understanding of diseases and how they affect overall brain connectivity.

\end{abstract}
\section{Introduction}
Neurodevelopmental conditions impair the growth and/or development of the brain. One such widely studied condition is autism spectrum disorder (ASD), which affects about 1.7\% of children in the US~\cite{baio_prevalence_2018}. ASD describes a spectrum of neurodevelopmental disorders characterized by atypical social behavior and sensory processing, where patients also demonstrate deficits in mental flexibility and high-level cognitive function~\cite{hong_atypical_2019,rudie_altered_2013}, and is currently diagnosed using cognitive assessment. Investigations suggest that ASD is a distributed disease and cannot be described by local effects, i.e. by specific brain regions, leading to an increased interest in applying connectomics for identifying differences in the autistic brain~\cite{hong_atypical_2019,muller_brain_2018}. 

Brain connectivity, or connectomics, and its topology has been widely studied, e.g., in healthy subjects~\cite{chung_classifying_2016,grayson_structural_2014,van_den_heuvel_rich-club_2011,zhao_age-related_2015,schirmer_structural_2018}, during early brain development~\cite{ball_rich-club_2014,schirmer_developing_2015,chung_characterising_2016}, and in disease~\cite{ktena_brain_2019,schirmer_rich-club_2019,collin_impaired_2013}. In particular, studies have used various ways to define subnetworks in the human connectome. These studies stratify groups of nodes in a brain network by a network theoretical measure, often relating it to the underlying network topology. Subsequent analyses often compare "traditional'' network measures between groups within the cohort or between subnetworks~\cite{schirmer_structural_2018,collin_impaired_2013,schirmer_network_2018,chung_concussion_rich-club_2019}. However, most subnetwork stratification, afer the brain network has been estimated, relies on a user-defined parameter, which can have significant impact on the subnetwork definition (e.g. $k$ in rich-club analyses~\cite{van_den_heuvel_rich-club_2011}). A recent study investigated the use of the network dependency index (NDI) to identify subnetworks in a data driven fashion and subsequently no user-parameter needs to be defined. In their study, Schirmer et al.~\cite{schirmer_network_2018} utilized structural connectomes with 170 brain regions in the NKI-Lifespan cohort to investigate the stability of NDI across age groups, and compared their method with subnetworks defined using the rich-club. NDI assigns a measure of "importance'' to each node in the connectome by quantifying the global effect of removing the node on network efficiency. Subsequent subnetwork stratification groups nodes automatically into Tiers based on this measure, identifying sets of nodes which can be considered essential for network efficiency. Importantly, their NDI framework demonstrated high reliability in determining a consistent set of regions to belong to the same subnetworks, without having to specify a user-defined parameter. However, they did not investigate the feasibility of applying their framework to functional data, or the utility of the identified subnetworks to identify group differences in a patient-control setting.

In this work, we apply the NDI framework to a set of resting-state functional connectomes in an ASD/control cohort based on the Automated Anatomical Labeling (AAL) atlas. We demonstrate that the framework can directly be applied to functional connectomes that are parcellated by a commonly used atlas on which few regions are defined. We also investigate the effect of using the cohort, control-only, and patient-only connectomes to derive NDI subnetworks. Additionally, we investigate the consistency of nodal subnetwork assignment by using different weighting schemes in the functional connectome, i.e. retaining only edges with negative weights, positive weights, and lastly the absolute weights of both, while varying the threshold for noise removal. Finally, we compare topological features in each of the subnetworks generated by all weighting schemes between subjects diagnosed with ASD and typically developing individuals. 

\section{Materials and Methods}
\subsection{Study design and patient population}
Data used in this study originates from the Autism Brain Imaging Data Exchange (ABIDE)~\cite{di_martino_autism_2014,craddock_neuro_2013} initiative and was downloaded through the python package \textit{nilearn}~\cite{nielsen_multisite_2013}. ABIDE consists of data comprising ASD (patients) and typically developing (controls) individuals~\cite{di_martino_autism_2014}. Each individual underwent a magnetic resonance imaging protocol, including rsfMRI and MPRAGE sequences. Details of acquisition, informed consent, and site-specific protocols are available elsewhere\footnote{\url{http://fcon_1000.projects.nitrc.org/indi/abide/}}. The cohort characteristics are summarized in Table~\ref{tab1}.

\begin{table}
\centering
\caption{Cohort characterization.}\label{tab1}
\begin{tabular}{|c|c|c|c|}
\hline
&  Cohort & Control & Patients\\
\hline
N & 819 & 440 & 379\\ \hline
Age, years, mean (sd) &  16.39 (7.12) & 16.27 (6.74) & 16.53 (7.54) \\
\hline
\end{tabular}
\end{table}

\subsection{RsfMRI preprocessing and group connectomes}
Data were preprocessed based on the ABIDE Connectome Computation System pipeline, which included slice timing and motion correction, removal of mean CSF and white matter signals, and detrending of linear and quadratic drifts. Subsequently, a temporal band-pass filtering was applied (0.01-0.1Hz) and the rsfMRI data registered to the MNI template. Regions for network analysis were defined based on the AAL atlas and the pre-processed time series was demeaned. Prior to network analysis, brainstem and cerebellar regions were removed, resulting in a total of 90 brain regions. Edge weights were computed as a covariance matrix~\cite{varoquaux_learning_2013} and edges with an absolute weight less than a given threshold were removed to reduce the effects of spurious signals. In this study, we investigate edge weight thresholds of 0.01, 0.03, and 0.05. In this study, we investigate three kinds of networks - by retaining only the positive weights (pos), only the absolute values of negative weights (neg), and the absolute of all weights (abs) - as there is no consensus on which of these are most discriminative. 

Subnetworks may be determined on group-averaged connectomes. Such connectomes have been used in multiple studies~\cite{van_den_heuvel_rich-club_2011,schirmer_structural_2018,schirmer_network_2018}. First, the binarized connectivity matrices of all subjects within a group are summarized by only retaining edges that are present in at least 90\% of the subjects (group adjacency matrix) with the goal to preserve connections which can be reliably identified. To allow for weighted network analyses, weights are added to the edges of the group adjacency matrix by averaging the edge weights from the contributing subjects' connectivity matrices. This process creates a weighted group-averaged connectome $W_{group}$, which can be utilized for analyses.

In this study, we group our cohort in three different ways. First, we create a \textit{cohort connectome}, which utilizes information from all subjects within the cohort. As our cohort contains patients and controls, we also generate a \textit{patient connectome}, as well as a \textit{control connectome} for NDI analysis.

\subsection{Network dependency index subnetworks}
A detailed description of the NDI framework is given elsewhere~\cite{schirmer_network_2018}. In brief, given a connectivity matrix $W_{group}=\{w_{ij}\}$ with $n$ nodes, we first calculate the full topological distance matrix $D$ between all node pairs based on the Dijkstra's algorithm and using the inverse of the connection strength $w_{ij}$ between nodes $i$ and $j$ as an initial topological distance. Subsequently, we derive the information measure $I_{ij}$ between nodes $i$ and $j$ given by $1/D_{ij}$. $I$ is then normalized by the maximum information measure. The NDI score of node $m$ is then given as 

\begin{eqnarray*}
	NDI_m &=& mean(\{I_i\}_{i=1,..,m-1,m+1,...,n} )\\
	 &=& mean( \{ \sum_j I_{ij} - (I_{-m})_{ij}\}_{i=1,..,m-1,m+1,...,n}  ), 
\end{eqnarray*}

where  $(I_{-m})_{ij}$ is the information measure of all nodes in the connectome from which node m has been removed. This analysis is then repeated for all nodes in the network, resulting in an nx1 dimensional feature vector of NDI scores for the network.

For each group connectome, we calculate its NDI scores. All nodes with an NDI of 0 are assigned to Tier 4. Using the natural-log-transformed NDI (excluding nodes with NDI=0), we apply a Gaussian Mixture Model with 3 Gaussian distributions (GMM), where subnetwork assignments are based on the halfway point between the Gaussian centers, resulting in three additional Tiers. In total, the nodes in a network are differentiated into four Tiers (including the NDI=0 Tier), where nodes within a Tier are ``similar'' with respect to their information measure.

\subsection{Network measures}
We utilize the subnetworks defined on the group connectome to stratify the nodes in each subject's connectome. Subsequently, we characterize each subnetwork's topology by calculating two commonly used network measures describing different aspects of the connectome organization, namely transitivity (T), and global efficiency (E)~\cite{rubinov_complex_2010}.

\subsection{Statistical analysis}
First, we compare the ranking of the regions in the brain, defined by NDI, between the different group connectomes. NDI assigns a measure of ``importance'' to each node in the connectome. According to this assignment, we can subsequently rank the regions in the brain. In order to compare different assignments from multiple group connectomes, we ultimately compare the resulting ranked lists. Here, we use the ranked-biased overlap (RBO) measure to estimate similarity of rankings, with higher weights for higher ranks~\cite{webber_similarity_2010}. In our setting this means that a variation in the order of important nodes is penalized more strongly, compared to the order among less important nodes. The closer the RBO value is to 1.0, the greater the agreement in node ranking between the three connectomes, with 1.0 representing complete similarity.

Subsequently, we calculate the network measures for each of the subnetworks defined on each group connectomes. This allows us to investigate topological differences between ASD patients and controls. We utilize the Mann-Whitney-Wilcox test and statistical significance was set at $p<0.05$. 

\section{Results}
\begin{figure}[ht!]
\centering
\includegraphics[width=.7\linewidth]{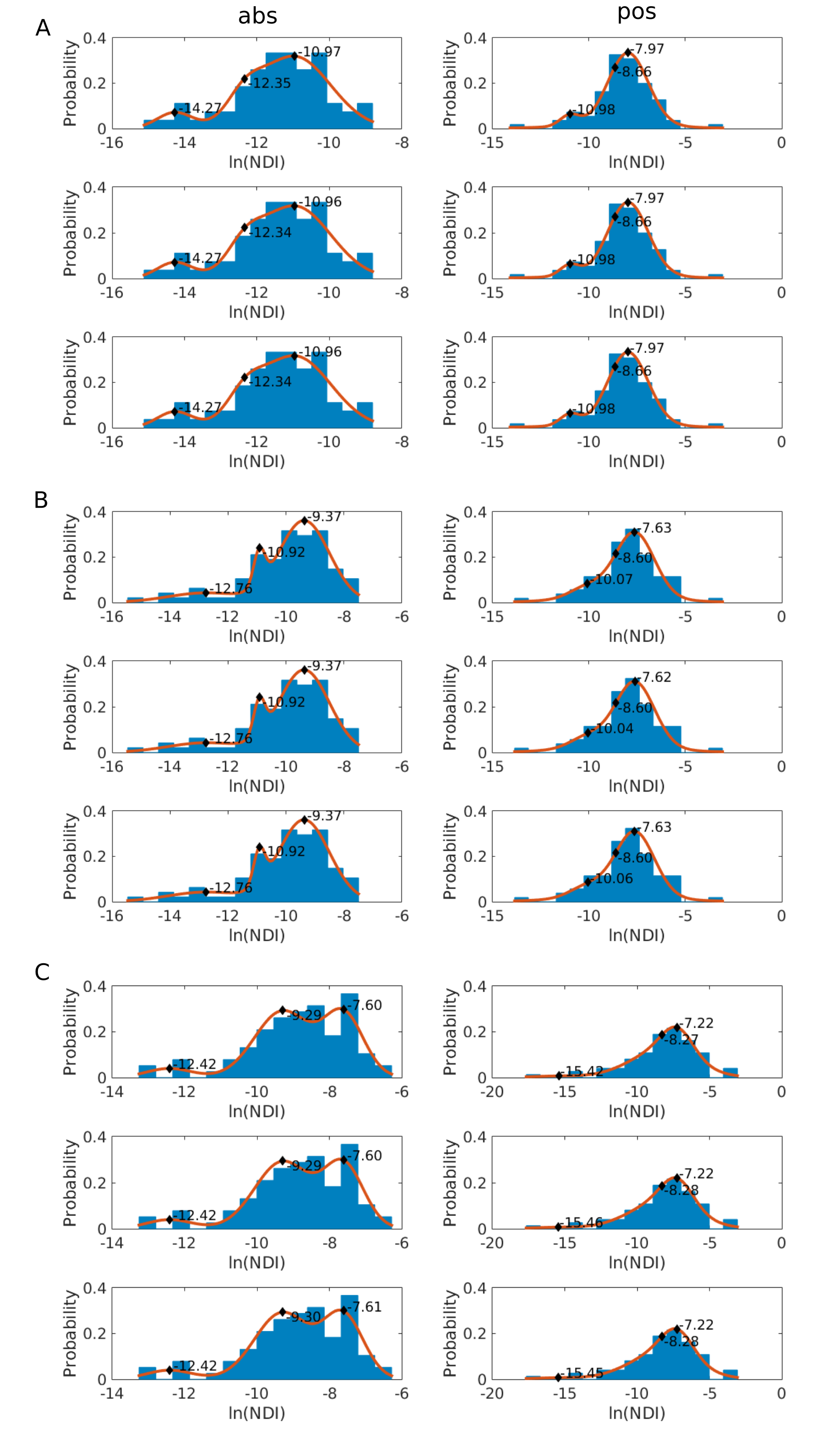}
\caption{Log-transformed NDI histograms for group connectomes based on absolute (abs; left column) and positive (pos; right column) edge weights. A, B, and C, correspond to edge thresholds of 0.01, 0.03, and 0.05, respectively. Each row of A, B, and C, corresponds (from top to bottom) to the cohort, patient-only, and healthy-only connectome. Centers of the three fitted Gaussians are indicated with a black diamond and the corresponding value is given to its right.}
\label{fig1}
\end{figure}

The log-transformed NDI scores with the fitted GMM model are shown in Figure~\ref{fig1} for each group connectome and for abs and pos weighting schemes. In case of retaining only negative weights, the population of calculated NDI value was so restricted, that no Gaussian fit was possible on neg matrices for all group connectomes. Therefore, for the remainder of this analysis, we restricted our analysis to abs and pos weightings only. In general, NDI scores are stable across group connectomes, with similar distributions and identified GMM centers within each threshold/weighting combination.

Table~\ref{tab2} summarizes the comparison of nodal ranking according to the NDI score. Firstly, for all combinations of threshold and weighting scheme, all three group connectomes resulted in differences of RBO values of less than 0.005. Secondly, RBO measures remained relatively consistent irrespective of threshold but differed more with weighting - meaning that edge thresholding has less effect on varying node rank across groups than weighting scheme. As node rankings were the same for all group connectomes for all threshold/weighting combination, we selected the Tier labels from the control connectome to stratify the nodes in each subject's network for the remainder of the analysis.

\begin{table}[ht!]
\setlength{\tabcolsep}{6pt}
  \centering
  \caption{Summary of RBO measures calculated on all group connectomes and for absolute (abs) and positive (pos) weights. Individual results of group connectomes are summarized by a single number, as they resulted in RBO differences of less than 0.005.}
    \begin{tabular}{cccccccc} 
          & \textbf{threshold} & \multicolumn{2}{c}{\textbf{0.01}} & \multicolumn{2}{c}{\textbf{0.03}} & \multicolumn{2}{c}{\textbf{0.05}} \\
    \cmidrule{2-8}  
    \multicolumn{1}{c}{\textbf{threshold}} & \textbf{weighting} & \multicolumn{1}{c}{\textbf{abs}} & \multicolumn{1}{c}{\textbf{pos}} & \multicolumn{1}{c}{\textbf{abs}} & \multicolumn{1}{c}{\textbf{pos}} & \multicolumn{1}{c}{\textbf{abs}} & \multicolumn{1}{c}{\textbf{pos}} \\
    \midrule
    \multirow{2}[4]{*}{\textbf{0.01}} & \textbf{abs} & 1.00     & 0.69  & 0.81  & 0.68  & 0.74  & 0.61 \\
\cmidrule{2-8}          & \textbf{pos} & 0.69  & 1.00     & 0.72  & 0.85  & 0.79  & 0.76 \\
    \midrule
    \multirow{2}[4]{*}{\textbf{0.03}} & \textbf{abs} & 0.81  & 0.72  & 1.00     & 0.68  & 0.81  & 0.62 \\
\cmidrule{2-8}          & \textbf{pos} & 0.68  & 0.85  & 0.68  & 1.00     & 0.73  & 0.78 \\
    \midrule
    \multirow{2}[4]{*}{\textbf{0.05}} & \textbf{abs} & 0.74  & 0.79  & 0.81  & 0.73  & 1.00     & 0.67 \\
\cmidrule{2-8}          & \textbf{pos} & 0.61  & 0.76  & 0.62  & 0.78  & 0.67  & 1.00 \\
    \end{tabular}%
  \label{tab2}%
\end{table}%

\begin{figure}[ht!]
\centering
\includegraphics[width=.7\linewidth]{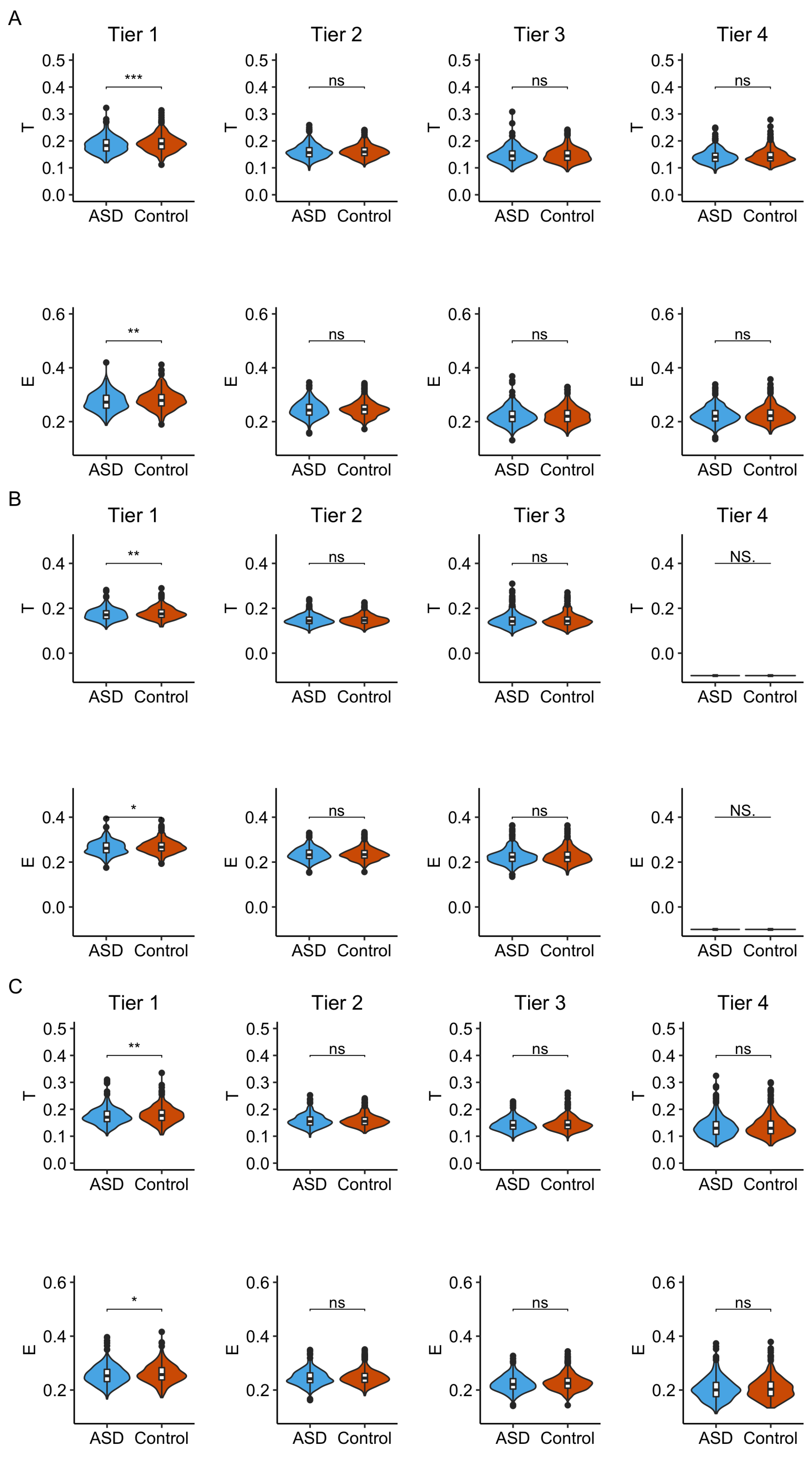}
\caption{Boxplot of topological network measures computed using absolute edge weights,
 for both ASD and Control groups from Tiers 1 to 4. A, B, and C correspond to edge thresholds of 0.01, 0.03, and 0.05, respectively. Statistical significance based on the Mann-Whitney-Wilcox test is indicated above each boxplot for transitivity (T) and efficiency (E). (ns: $p>0.05$; *: $p<0.05$; **: $p<0.01$; ***: $p<0.001$)}
\label{fig2}
\end{figure}

\begin{figure}[ht!]
\centering
\includegraphics[width=.7\linewidth]{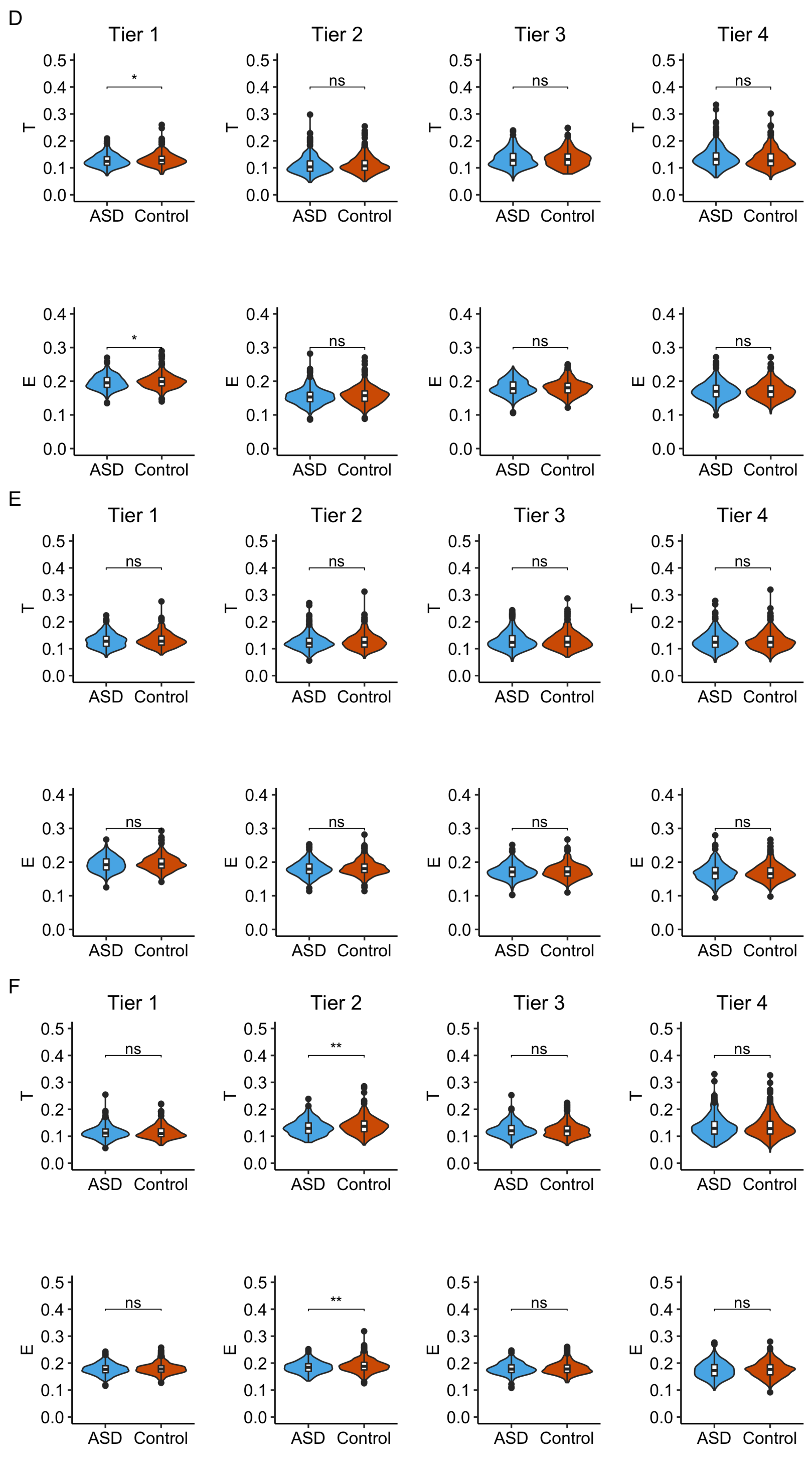}
\caption{Boxplot of topological network measures computed using positive edge weights,
 for both ASD and Control groups from Tiers 1 to 4. D, E, and F correspond to edge thresholds of 0.01, 0.03, and 0.05, respectively. Statistical significance based on the Mann-Whitney-Wilcox test is indicated above each boxplot for transitivity (T) and efficiency (E). (ns: $p>0.05$; *: $p<0.05$; **: $p<0.01$; ***: $p<0.001$)}
\label{fig3}
\end{figure}

Figures~\ref{fig2} and~\ref{fig3} compare network topological measures in each of the subnetworks between ASD and controls, for absolute and positive weighting schemes, respectively. In the case of an edge threshold of 0.03 on the abs weighting, only one node had a value of NDI=0, which meant that the network measures computed were ill-defined on this subnetwork (Tier 4). We observe significant differences in both T and E in Tier 1 in the case of using the absolute weighting for all thresholds, and in Tier 1 and Tier 2 in case of positive weighting, for a threshold of 0.01 and 0.05, respectively. 

For absolute weights only Tier 1 showed significant differences. Regions which were consistently identified as Tier 1 regions across thresholds were the precentral gyrus, median cingulate and paracingulate gyri, cuneous, precuneous, superior occipital gyrus, fusiform gyrus,  and the supramarginal gyrus, all in the right hemisphere. Left hemisphere Tier 1 regions consisted of the orbital part of inferior frontal gyrus and medial superior frontal gyrus. Only the insula was identified bilaterally. For positive weights, we observed significant differences in Tier 1 and Tier 2 for a threshold of 0.01 and 0.05, respectively. Regions that were consistently identified in both tiers were the orbital part of middle frontal gyrus, median cingulate and paracingulate gyri, and caudate. In the left hemisphere, the regions consisted of the superior and inferior parietal gyrus, and the putamen. Additionally, the triangular part of inferior frontal gyrus was identified in both hemispheres. 

\section{Discussion}
In this study, we showed that the NDI framework for defining subnetworks can be applied to resting-state functional connectomes and demonstrated its consistency regardless of the group connectome used. Additionally, we demonstrated that there are topological group differences in the subnetworks generated, when comparing subjects diagnosed with ASD and typically developing individuals. 

Applying NDI for subnetwork definition worked for functional networks, utilizing an atlas with approximately half the number of regions, compared to the original study (which employed the Craddock200 atlas with 170 regions~\cite{schirmer_network_2018}). While it is possible that a reduction in the number of connectome regions can increase the variation in the GMM fitted Gaussian centers, we observed stable estimations for these centers for our three ABIDE group connectomes. In addition, we showed better agreement of nodal assignment, if weighting scheme is held constant while varying the threshold, compared to varying weighting scheme while holding the threshold constant. This further highlights the stability for the reliable estimation of subnetworks using NDI. Importantly, we showed that by stratifying the individual connectomes by subnetworks, we were able to find significant group differences between individuals with ASD and controls in regions belonging to Tier 1 or Tier 2, i.e. regions which are more important for efficient information transport. The only region consistently identified across thresholds and weighting schemes was the right median cingulate and paracingulate gyri, which was recently highlighted in ABIDE using a neural network approach~\cite{bi2018diagnosis}. Subsequent analyses may use this information, along with other regions identified from the appropriate weighting scheme, as priors when aiming to investigate specific regions

There are general network analysis limitations, which may further affect our study. Although it is quite common to threshold functional connectivity matrices in order to reduce the effect of noise, there is no agreed upon method on how to define this threshold. Therefore, studies commonly investigate different thresholds with the aim to demonstrate consistency of results. Following this reasoning, we investigated a variety of methods of thresholding, specifically by using only the positive and only the negative edge weights, as well as the absolute edge weight, and by thresholding each at levels of 0.01, 0.03, and 0.05. While we show that the framework can still be utilized, except in the case of negative-only weights, there are many more thresholds which can be investigated. This will be the aim of future work. In our study, we analyze connectomes with 90 regions, whereas the original publication was able to utilize 170. While the appropriate atlas, or number of regions in the brain, remains an open question, a larger number of regions results in more data on which the three Gaussians can be estimated. In future work we aim to investigate agreement of NDI based Tier-assignment by utilizing multiple atlases, and mapping our results back to the brain template to identify spatial patterns of the regions in each Tier. In this work we estimated NDI subnetwork definition based on average group connectomes.  However, it is possible to use the connectomes of each individual subjects to determine the subnetworks, which may help to further differentiate subtypes of diseases. While this is an interesting objective for future work, the primary aim here was to demonstrate that the NDI framework can be utilized in functional data and that it can identify group differences in case of disease. 

In conclusion, we demonstrated that the NDI subnetwork framework can be applied to functional connectomes and produces stable results, when modifying the population from which the group connectome is generated (patients versus control). In addition, we show that these subnetwork definitions can be utilized to show group differences between individuals diagnosed with ASD and healthy controls, where those differences are mainly located in nodes/brain regions with highest importance.

\subsection*{Acknowledgments}
This project has received funding from the European Union's Horizon 2020 research and innovation programme under the Marie Sklodowska-Curie grant agreement No 753896 (MDS) and the American Heart Association and Children's Heart Foundation Postdoctoral Fellowship, 19POST34380005 (AWC).

\bibliographystyle{splncs04}

\end{document}